\documentclass[aps, prl, twocolumn, superscriptaddress]{revtex4}
\usepackage{graphicx}
\usepackage{array}
\usepackage{amsmath,amssymb}
\usepackage[dvipsnames,usenames]{color}
\begin{document}

\title{Nonequilibrium Kondo effect by equilibrium numerical renormalization group method: The hybrid Anderson model subject to a finite spin bias}
\author{Tie-Feng Fang}
\email{fangtiefeng@lzu.edu.cn}
\affiliation{Center for Interdisciplinary Studies, School of Physics, Lanzhou University, Lanzhou 730000, China}
\author{Ai-Min Guo}
\affiliation{Hunan Key Laboratory for Super-microstructure and Ultrafast Process, School of Physics and Electronics, Central South University, Changsha 410083, China}
\author{Qing-Feng Sun}
\email{sunqf@pku.edu.cn}
\affiliation{International Center for Quantum Materials, School of Physics, Peking University, Beijing 100871, China}
\affiliation{Collaborative Innovation Center of Quantum Matter, Beijing 100871, China}
\affiliation{CAS Center for Excellence in Topological Quantum Computation, University of Chinese Academy of Sciences, Beijing 100190, China}
\date{\today}
\begin{abstract}
We investigate Kondo correlations in a quantum dot with normal and superconducting electrodes, where a spin bias voltage is applied across the device and the local interaction $U$ is either attractive or repulsive. When the spin current is blockaded in the large-gap regime, this nonequilibrium strongly-correlated problem maps into an equilibrium model solvable by the numerical renormalization group method. The Kondo spectra with characteristic splitting due to the nonequilibrium spin accumulation are thus obtained at high precision. It is shown that while the bias-induced decoherence of the spin Kondo effect is partially compensated by the superconductivity, the charge Kondo effect is enhanced out of equilibrium and undergoes an additional splitting by the superconducting proximity effect, yielding four Kondo peaks in the local spectral density. In the charge Kondo regime, we find a universal scaling of charge conductance in this hybrid device under different spin biases. The universal conductance as a function of the coupling to the superconducting lead is peaked at and hence directly measures the Kondo temperature. Our results are of direct relevance to recent experiments realizing negative-$U$ charge Kondo effect in hybrid oxide quantum dots [Nat.\,\,Commun.\,\,\textbf{8}, 395 (2017)].
\end{abstract}
\maketitle

\section{I. introduction}
The Kondo effect \cite{Hewson1993}, a paradigm of strongly correlated physics, has been revived for two decades in artificial nanostructures such as quantum dots (QDs) \cite{Goldhaber-Gordon1998, Cronenwett1998}. It describes the many-body screening of a local spin by conduction electrons. Unlike bulk materials, nanoscale Kondo systems are routinely driven out of equilibrium by applying charge or spin bias voltages \cite{Kobayashi2010} across the devices. In such a nonequilibrium situation, the local spin is exposed to different Fermi levels, which opens up inelastic channels and deeply influences the many-body correlations. Although the voltage splitting of the Kondo resonance can be roughly captured in various perturbative calculations \cite{Meir1993, Wingreen1994, Rosch2001, Lebanon2001, Fujii2003, Shah2006, Fritsch2010, Nuss2012}, precisely describing the Kondo effect out of equilibrium and its decoherence caused by the bias and the current is a long-standing challenge even in a steady state.

Specifically, exact solutions \cite{Schiller1995, Katsura2007, Breyel2011, Posske2013, Bolech2016, Beri2017} at the Toulouse point of nonequilibrium Kondo models are inapplicable to the more microscopic Anderson model. Quantum Monte Carlo \cite{Han2007, Muhlbacher2011, Gull2011, Cohen2014} and master equation \cite{Scully1997, Dorda2014, Dorda2015, Titvinidze2015, Dorda2016} methods cannot access the strong-coupling Kondo limit at zero temperature. An exciting prospect of studying nonequilibrium steady states of correlated nanostructures is offered by the scattering-states numerical renormalization group (NRG) \cite{Anders2008, Schmitt2011} combined with the time-dependent NRG \cite{Anders2005, Eidelstein2012, Nghiem2014, Nghiem2017}. But the incomplete thermalization at long times still remains an issue \cite{Nghiem2014, Nghiem2017, Rosch2012}. In view of these difficulties, it is helpful to find some special steady-state nonequilibrium situations that can be transformed into equilibrium ones. The related nonequilibrium Kondo problem can then be solved by the most powerful NRG method at equilibrium \cite{Wilson1975, Krishna-murthy1980, Bulla2008}, thereby yielding quantitative insight on the many-body correlations out of equilibrium and providing useful benchmarks for the development of truly nonequilibrium methods.

Despite the theoretical challenge, recent advances in nanofabrication techniques continue to produce exotic Kondo systems. QDs with local attraction ($U<0$) have now been fabricated both at the $\textrm{LaAlO}_3/\textrm{SrTiO}_3$ interface \cite{Cheng2015, Cheng2016, Tomczyk2016, Prawiroatmodjo2017} and in carbon nanotubes \cite{Hamo2016}. Such QD devices can support a charge Kondo effect instead of the conventional positive-$U$ spin Kondo effect. In bulk materials, the negative-$U$ charge Kondo effect was first proposed \cite{Taraphder1991} and then realized \cite{Costi2012, Matsushita2005, Matusiak2009, Dzero2005} long ago. Only very recently has this charge Kondo effect been demonstrated in highly tunable QDs \cite{Prawiroatmodjo2017}, yielding transport characteristics consistent with previous theoretical predictions \cite{Andergassen2011, Cheng2013, Koch2007, Zitko2006, Cornaglia2004, Fang2014}. While all these theories concerned only coherent coupling to normal-state leads, the experiments \cite{Cheng2015, Cheng2016, Tomczyk2016, Prawiroatmodjo2017} have exhibited the great flexibility on coupling geometries, where the two leads coupled to the negative-$U$ QD can be tuned between the superconducting (S) and normal (N) states. This offers unique opportunities to study the interplay of charge Kondo and superconducting correlations. We recently showed \cite{Fang2017} that Cooper-pair tunneling processes in a S-QD-S device act as a transverse pseudo magnetic field to the charge Kondo effect, but the intriguing Kondo splitting by this field is invisible in the gapped density of states. Here, we circumvent this problem by exploiting a hybrid N-QD-S geometry where the energy scales of the charge Kondo effect and the pseudofield can be tuned independently.

The hybrid geometry also provides a realization of the desired nonequilibrium steady state: our N-QD-S device driven nonequilibrium by a spin bias voltage maps into an equilibrium model, when the spin current is fully blockaded by superconducting pairing. The key ingredient of the mapping is a canonical transformation which transforms the nonequilibrium spin accumulation in the N lead into a magnetic field acting on the dot. By studying such a hybrid system, intriguing nonequilibrium features of the negative-$U$ charge Kondo effect, distinctive from those of the positive-$U$ spin Kondo effect, can be quantitatively addressed. Note that in normal systems the spin Kondo effect under a spin bias was already observed \cite{Kobayashi2010, Taniyama2003, Hamaya2016}, but only described by the crude equation-of-motion (EOM) approach \cite{Qi2008, Lim2013, Sahoo2016}.

In this paper, we present a quantitative study of these nonequilibrium strongly correlated effects by using the equilibrium NRG method. Characteristic Kondo spectra with spin-bias-induced splitting are accurately calculated. We find that the charge Kondo effect is enhanced out of equilibrium. It undergoes an additional splitting by the superconducting proximity effect, giving rise to four Kondo peaks in the local density of states and two peaks in the linear charge conductance as a function of the spin bias. Interestingly, the conductance as a function of the coupling to the S lead exhibits a universal scaling which yields a direct measurement of the nonequilibrium Kondo temperature. These intriguing features might be verified in $\textrm{LaAlO}_3/\textrm{SrTiO}_3$-based QDs \cite{Cheng2015, Cheng2016, Tomczyk2016, Prawiroatmodjo2017} where the negative-$U$ charge Kondo effect has already been observed \cite{Prawiroatmodjo2017}. By contrast, the spin Kondo effect, which is decohered by the spin bias and partially restored by the superconductivity, has no significant transport consequences.

The rest of the paper is organized as follows. In Sec.\,II, we introduce the model Hamiltonian and provide some necessary details of our theoretical method. Numerical results and their discussion are presented in Sec.\,III. Finally, Sec.\,IV is devoted to a summary.

\section{II. model and method}
Our N-QD-S device is modeled by the hybrid Anderson Hamiltonian:
\begin{eqnarray}
H&=&H_D+\sum_{L=N,S}(H_L+H_{L D}),\\
H_D&=&\sum_\sigma\varepsilon_{\textrm{d}}d^\dagger_\sigma d_\sigma+Un_\uparrow n_\downarrow,\\
H_L&=&\sum_{k,\sigma}(\varepsilon_{k}+\delta_{\scriptscriptstyle{L N}}\mu_{\scriptscriptstyle{N}\scriptstyle{\sigma}})C^\dagger_{L k\sigma}C_{L k\sigma}\nonumber\\
&&\,+\,\,\,\delta_{\scriptscriptstyle{L S}}\sum_{k}(\Delta C^\dagger_{S k\uparrow}C^\dagger_{S\bar{k}\downarrow}+\textrm{H.c.}),\\
H_{LD}&=&\sum_{k,\sigma}V_L d^\dagger_\sigma C_{L k\sigma}+\textrm{H.c.}
\end{eqnarray}
Here $H_D$ models the isolated dot in which the operator $d_\sigma$ ($n_\sigma\hspace{-0.3mm}=\hspace{-0.3mm}d^\dagger_\sigma d_\sigma$) annihilates an electron of energy $\varepsilon_{\textrm{d}}$ and spin $\sigma\hspace{-0.7mm}=\,\uparrow,\,\downarrow$. The onsite interaction $U$ is either attractive ($U<0$) or repulsive ($U>0$). $H_L$ describes the normal ($L=N$) lead with its spin-dependent chemical potential $\mu_{\scriptscriptstyle{N}\scriptstyle{\sigma}}$ driven by the charge $W\equiv\frac{1}{2}(\mu_{\scriptscriptstyle{N}\uparrow}+\mu_{\scriptscriptstyle{N}\downarrow})$ and spin $V\equiv\frac{1}{2}(\mu_{\scriptscriptstyle{N}\uparrow}-\mu_{\scriptscriptstyle{N}\downarrow})$ biases, or the superconducting ($L=S$) lead with chemical potential $\mu_{\scriptscriptstyle{S}}=0$ and an energy gap $\Delta$. The operator $C_{L k\sigma}$ annihilates an electron of wave vector $k$ ($\bar{k}\hspace{-0.2mm}=\hspace{-0.2mm}-k$) and energy $\varepsilon_k$ in lead $L$. $H_{LD}$ represents the dot-lead tunneling characterized by the amplitudes $V_L$, which define two tunneling rates: $\Gamma=\pi\rho V_N^2$ at the N-QD interface and $\Gamma_S=\pi\rho V_S^2$ at the QD-S interface, with $\rho$ the lead density of states.

The difficulty to obtain the nonequilibrium steady-state properties of $H$ lies in that the density operator $\rho_{\scriptscriptstyle{H}}$ is not explicitly known for finite bias. We can, however, eliminate the potential difference of the N and S leads by a time-dependent canonical transformation:
\begin{equation}
H'(t)=\mathcal{U}(t)H\mathcal{U}^\dagger(t)+i\hbar\dot{\mathcal{U}}(t)\mathcal{U}^\dagger(t),
\end{equation}
with the unitary operator $\mathcal{U}(t)$ given by
\begin{equation}
\mathcal{U}(t)=\textrm{exp}\bigg[\frac{it}{\hbar}\sum_{k,\sigma}\mu_{\scriptscriptstyle{N}\scriptstyle{\sigma}}(C^\dagger_{Nk\sigma}C_{Nk\sigma}+d^\dagger_\sigma d_\sigma)\bigg].
\end{equation}
Under this transformation, the operators $d_\sigma$, $C_{Nk\sigma}$, and $C_{Sk\sigma}$ become
\begin{eqnarray}
\mathcal{U}(t)d_\sigma\mathcal{U}^\dagger(t)&=&d_\sigma\exp\hspace{-0.5mm}\big[\hspace{-0.5mm}-\hspace{-0.5mm}(it/\hbar) \mu_{\scriptscriptstyle{N}\scriptstyle{\sigma}}\big],\\
\mathcal{U}(t)C_{Nk\sigma}\mathcal{U}^\dagger(t)&=&C_{Nk\sigma}\exp\hspace{-0.5mm}\big[\hspace{-0.5mm}-\hspace{-0.5mm}(it/\hbar) \mu_{\scriptscriptstyle{N}\scriptstyle{\sigma}}\big],\\
\mathcal{U}(t)C_{Sk\sigma}\mathcal{U}^\dagger(t)&=&C_{Sk\sigma}.
\end{eqnarray}
The transformed Hamiltonian can be written as
\begin{equation}
H'(t)=H'_D+H'_N+H_{ND}+H_S+H'_{SD}(t),
\end{equation}
where $H'_D$ differs from $H_D$ by $\varepsilon_{\textrm{d}}\rightarrow\varepsilon_{\textrm{d}}-W-\sigma V$, $H'_N$ differs from $H_N$ by $\mu_{\scriptscriptstyle{N}\scriptstyle{\sigma}}\rightarrow0$, and $H'_{SD}(t)$ differs from $H_{SD}$ by $V_S\rightarrow e^{\frac{it}{\hbar}(W+\sigma V)}V_S$. Hence the two leads are both held at zero chemical potential. But the difficulty remains since $H'(t)$ is now time dependent. In the superconducting limit where the gap $\Delta$ is the largest energy scale except the bandwidth, the Hamiltonian term $H_S+H'_{SD}(t)$ exactly reduces to \cite{Fang2017} $H_{DD}=\Gamma_Sd^\dagger_\uparrow d^\dagger_\downarrow+\textrm{H.c.}$ for vanishing charge bias $W$. As a result,
\begin{equation}
H'=H'_D+H'_N+H_{ND}+H_{DD}
\end{equation}
becomes time independent, reaching the desired nonequilibrium to equilibrium mapping. This procedure transforms the nonequilibrium spin accumulation in the N lead into a magnetic field acting on the dot and yields an equilibrium density operator $\rho_{\scriptscriptstyle{H'}}=e^{-\beta H'}\hspace{-1mm}/\hspace{0.5mm}\textrm{Tr}e^{-\beta H'}$. The physical implication is that the N-QD-S system driven by the spin bias $V$ is somewhat equivalent to an equilibrium model when the spin current is fully blockaded by the S lead. For finite $\Delta$ comparable to other energy scales, a nonzero spin current can exist, which violates the mapping.

We first calculate the equilibrium properties of $H'$ by using the highly accurate NRG method \cite{Wilson1975, Krishna-murthy1980, Bulla2008} based on the full density matrix algorithm \cite{Anders2005, Peters2006, Weichselbaum2007, Fang2015}. Then an inverse transformation is performed to obtain the nonequilibrium properties of the original Hamiltonian $H$. For example, in the superconducting limit and at $W=0$, the retarded Green's functions $G_{d_\sigma,B}(\varepsilon)\equiv\langle\langle d_\sigma|B\rangle\rangle_H$ of the original Hamiltonian $H$ and $ G'_{d_\sigma,B}(\varepsilon)\equiv\langle\langle d_\sigma|B\rangle\rangle_{H'}$ of the transformed Hamiltonian $H'$ are related by
\begin{equation}
G_{d_\sigma,B}(\varepsilon)=G'_{d_\sigma,B}(\varepsilon-\sigma V),
\end{equation}
where $B$ stands for arbitrary operators. NRG calculations are performed by using a discretization parameter $\Lambda=1.8$ for dynamical properties and $\Lambda=4$ for thermodynamic quantities, and retaining $M_K=1200\hspace{-0.3mm}\sim\hspace{-0.5mm}1600$ states per iteration. Discrete spectral data is smoothened based on the log-Gaussian kernel proposed in Ref.\,\cite{Weichselbaum2007} with a broadening parameter $\alpha=0.3$. Results are z averaged over $N_z=4$ calculations.

\section{III. results and discussions}
What follows are numerical results at zero temperature. We fix $|U|=15\Gamma=0.012D$, $2\varepsilon_\textrm{d}+U=0$, and the half bandwidth $D=10$, resulting in a Kondo temperature \cite{Hewson1993} $T_K=\sqrt{|U|\Gamma/2}\,\textrm{exp}(-\pi|U|/8\Gamma)\simeq6.06\times10^{-5}\simeq7.57\times10^{-3}\Gamma$ due to cotunneling processes at the N-QD interface. Note that $T_K$ is the common energy scale of the charge and spin Kondo effects at equilibrium.

Figure 1 compares the nonequilibrium charge and spin Kondo effects in the N-QD subsystem when the QD-S coupling $\Gamma_S=0$. It is shown that negative-$U$ charge Kondo correlations are robust against the nonequilibrium driven by the spin bias $V$. The bias can split but never suppress the charge Kondo resonance in the local density of states $A(\varepsilon)\equiv-\frac{1}{\pi}\sum_\sigma\textrm{Im}\langle\langle d_\sigma|d^\dagger_\sigma\rangle\rangle_H$, as demonstrated in Figs.\,1(a) and 1(b). In particular, the split sharp Kondo peaks, one at each spin-resolved chemical potential, persist to very large bias of $V\hspace{-1mm}\gg\hspace{-1mm}T_K$, until they merge into the Hubbard bands for $V$ approaching $|U|/2$ where pseudospin-flip scattering and hybridization processes are indistinguishable. The robustness is expected since the two-electron cotunneling processes composing the charge Kondo effect are elastic for arbitrary $V$ [Fig.\,1(d)], i.e., the spin voltage cannot decohere Kondo correlations. Accordingly, the local pseudospin \cite{Fang2017} is always fully screened: $Q^+\hspace{-0.5mm}=\hspace{-0.5mm}Q^-\hspace{-0.5mm}=\hspace{-0.5mm}\langle d^\dagger_\downarrow d^\dagger_\uparrow\rangle\hspace{-0.5mm}=\hspace{-0.5mm}0$, $Q_z\hspace{-0.5mm}=\hspace{-0.5mm}\frac{1}{2}(n-1)\hspace{-0.5mm}=\hspace{-0.5mm}0$. Note that in the Hamiltonian $H'$ the spin voltage acts as a magnetic field whose influence on the charge Kondo effect is equivalent to a gate voltage in the spin Kondo effect. This implies a characteristic energy scale \cite{Hewson1993} $T^V_K=\sqrt{|U|\Gamma/2}\,\textrm{exp}\big(\pi\frac{V^2-U^2/4}{2\Gamma|U|}\big)$ for the nonequilibrium charge Kondo effect in the original Hamiltonian $H$. Since $T^V_K$ increases with $|V|$, the charge Kondo effect is actually enhanced out of equilibrium.

\begin{figure}[]
\includegraphics[width=1.0\columnwidth]{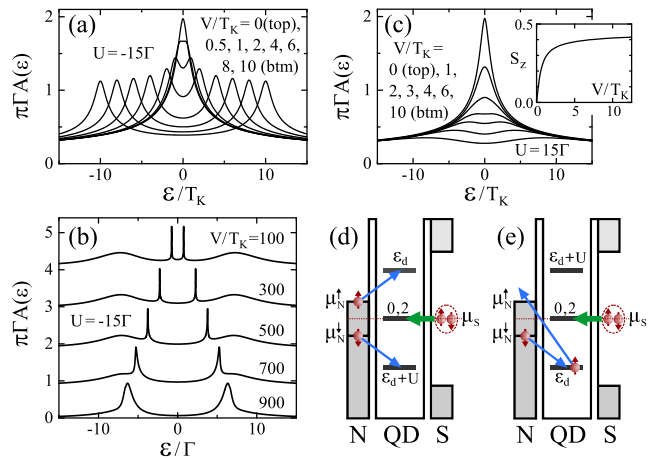}
\caption{Local spectral function $A(\varepsilon)$ in the nonequilibrium charge (a),(b) and spin (c) Kondo effects driven by various spin voltages $V$ for vanishing QD-S coupling. Inset of (c): Expectation value of the z-axis spin, $S_z$, of the spin Kondo QD as a function of $V$. (d) Schematic of the two-electron Kondo cotunneling process at the N-QD interface and the Cooper-pair tunneling process at the QD-S interface for finite $V$ and $U<0$. (e) Same as (d) but for $U>0$. Curves in (b) are offset for clarity.}
\end{figure}

It is not the case for the positive-$U$ spin Kondo effect. As shown in Fig.\,1(c), the spin Kondo resonance in $A(\varepsilon)$, which also splits into two peaks near $\varepsilon=\pm V$, rapidly fades with increasing the bias $V$. Accordingly, the local spin is no longer fully screened: $S_z=\frac{1}{2}(n_\uparrow-n_\downarrow)$ increases from $0$ [inset of Fig.\,1(c)], while $S^+=S^-=\langle d^\dagger_\uparrow d_\downarrow\rangle=0$. These are manifestations of nonequilibrium decoherence of the spin Kondo singlet. The underlying spin-flip cotunneling processes are inelastic and incur an energy cost of $2V$ [Fig.\,1(e)], which is the main source of the decoherence. Previous EOM study \cite{Qi2008} of the spin Kondo effect for magnetic impurities in nonmagnetic conductor driven by a spin bias (equivalent to our N-QD subsystem) claimed that the amplitude of the split Kondo peaks remains robust against the bias and there is no decoherence effect. Apparently, our accurate NRG data
invalidates the crude EOM results.

We now turn to the influence of the S lead by switching on the QD-S coupling $\Gamma_S$. Cooper-pair tunneling processes are thus allowed at the QD-S interface [Figs.\,1(d) and 1(e)], which are described by the local pairing term $H_{DD}$ arising from the superconducting proximity effect. For the charge Kondo effect, $H_{DD}$ serves as a transverse pseudo magnetic field applied on the QD. We emphasize that a real magnetic field in arbitrary directions does not split the charge Kondo resonance in the density of states, nor does a longitudinal pseudo magnetic field \cite{Andergassen2011, Fang2017}. But the transverse pseudo magnetic field discussed here, i.e., the QD-S coupling $\Gamma_S$, will cause such a splitting. Specifically, each of the two charge Kondo peaks in the nonequilibrium $A(\varepsilon)$, at each spin-resolved chemical potential of the N lead, splits into two peaks when $\Gamma_S$ increases exceeding $T^V_K$, giving rise to four peaks at $\varepsilon=\pm V\pm 2\Gamma_S$ [Fig.\,2(a)]. Detailed evolutions of these four Kondo peaks, including the two inner peaks merging at $\Gamma_S=V/2$ and re-splitting for $\Gamma_S>V/2$, are presented in Fig.\,2(b). Unlike the nonequilibrium splitting which is
coherent to charge Kondo correlations, the Cooper-pair tunneling at the QD-S interface is a decoherence process that disturbs the coherent superposition of all the two-electron cotunneling events in the charge Kondo state forming at the N-QD interface. Therefore, in addition to altering the energies of the four Kondo peaks, progressively stronger QD-S coupling also acts to suppress their amplitudes and eventually eliminates all the Kondo spectral weights. Meanwhile, the local pseudospin is transversely polarized by this pseudofield [Fig.\,2(d)].

\begin{figure}[]
\includegraphics[width=1.0\columnwidth]{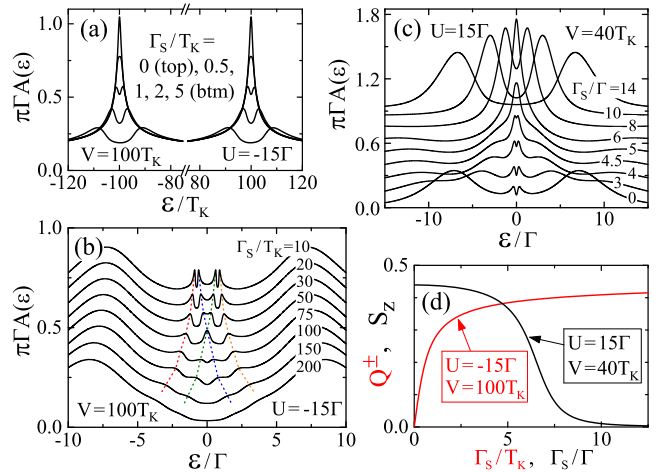}
\caption{Nonequilibrium spectral function $A(\varepsilon)$ under finite spin voltage $V$ in the negative-$U$ charge Kondo regime (a),(b) and the positive-$U$ regime (c) for different QD-S couplings $\Gamma_S$. (d) Expectation values of the transverse pseudospin, $Q^\pm$, of the attractive QD (red) and the longitudinal spin, $S_z$, of the repulsive QD (black) as functions of $\Gamma_S$, under finite $V$. Curves in (b),(c) are offset for clarity and dashed lines in (b) are guides for the evolutions of Kondo peaks.}
\end{figure}

On the other hand, the S lead influences the spin Kondo effect in a strikingly different way. It is known \cite{Deacon2010, Hofstetter2010, Martin-Rodero2011, Yamada2011, Koga2013, Baranski2013, Lee2014, Zitko2015, Weymann2015, Domanski2016, Sun2001, Cao2017} that at equilibrium the increasing $\Gamma_S$ can drive a crossover from the spin Kondo singlet to the BCS singlet, featuring characteristic spectral evolution as follows. Two Andreev resonances, which are split from the Hubbard bands at $\Gamma_S=0$, move toward the Fermi energy and merge with the Kondo peak to produce a single resonance at $\varepsilon=0$. This resonance splits again by further increasing $\Gamma_S$, signaling the end of the crossover into the BCS phase. These features also appear in the nonequilibrium $A(\varepsilon)$, as shown in Fig.\,2(c), where the crossover is studied under a finite spin bias and thus the Kondo peak is split. Remarkably, before it merges with the Andreev peaks, as $\Gamma_S$ increases, the nonequilibrium Kondo resonance is enhanced and its splitting shrinks [see the curves with $\frac{\Gamma_S}{\Gamma}=0\hspace{-0.5mm}\sim\hspace{-0.5mm}5$ in Fig.\,2(c)], due to the occurrence of excess Andreev-normal cotunneling \cite{Sun2001, Cao2017, Note1} in this regime. Coherent superposition of this Andreev-normal cotunneling process and the conventional two-electron Kondo cotunneling at the N-QD interface results in the enhancement of the spin Kondo effect. This effectively diminishes the spin voltage relative to the Kondo temperature, thereby shrinking the Kondo peak splitting and slowly decreasing the longitudinal polarization [see the $S_z$ curve in $0<\frac{\Gamma_S}{\Gamma}<5$ in Fig.\,2(d)]. Further increasing $\Gamma_S$ leads to the even dot occupancy, which is characterized by a rapid reduction of $S_z$ for $\frac{\Gamma_S}{\Gamma}>5$ in Fig.\,2(d).

Interestingly, these distinctive spectral features between the attractive and repulsive QDs are reflected in transport properties. While the spin current is blockaded in our device under arbitrary spin bias $V$, the charge bias $W$ can always drive a charge current $I$. The linear charge conductance $G\equiv\frac{\textrm{d}I}{\textrm{d}W}\big|_{W=0}$ is determined by the system with vanishing $W$. Therefore, our nonequilibrium-to-equilibrium mapping remains valid. At zero temperature, the conductance is \cite{Martin-Rodero2011,Tanaka2007,Sun1999}
$G=\frac{8e^2}{h}\Gamma^2\big[\big|G_{d_\uparrow,d_\downarrow}(V)\big|^2+\big|G_{d_\downarrow,d_\uparrow}(-V)\big|^2\big]$.
To facilitate NRG calculations of the Green's functions at the Fermi energy after the mapping, we have followed Ref.\,\cite{Tanaka2007} to apply the Bogoliubov transformation and use the Fermi-liquid relations. Results are given in Fig.\,3.

It is emphasized that the intriguing splitting of the negative-$U$ charge Kondo resonance by the superconducting proximity effect also shows up in the charge conductance as a function of the spin bias. As illustrated in Fig.\,3(a), on increasing the coupling to the S lead, the ``zero-bias anomaly'' in the conductance increases up to the unitary limit for $\Gamma_S<T_K$ and then splits for $\Gamma_S>T_K$. The split conductance peaks always reaching the unitary limit do not fade with increasing $\Gamma_S$. We elaborate the underlying physics as follows. In the charge Kondo regime, the current through the N-QD-S device is mediated by the charge Kondo cotunneling at the N-QD interface, with the many-body tunneling rate given by the Kondo temperature $T^V_K$, and the Cooper-pair tunneling at the QD-S interface, with the tunneling rate $\Gamma_S$. These two tunneling processes are compatible in the sense that they are confined in the same even-occupied subspace, both causing the QD to fluctuate between the degenerate $n=0,\,2$ states. Therefore, the unitary limit of the conductance is reached whenever $\Gamma_S$ and $V$ satisfy the condition $\Gamma_S=\alpha T^V_K$ of identical tunneling rates at the two interfaces. Here the constant $\alpha$, whose value will be given later, is of the order of $1$ but does not exactly equal $1$ because the Kondo temperature $T^V_K$ is not a well-defined energy scale. $T^V_K$ can differ from the true many-body tunneling rate at the N-QD interface by a constant multiplicative factor. Tuning $\Gamma_S$ and/or $V$ away from this condition results in the conductance decreasing steadily from the unitary limit. This scenario explains the evolutions of conductance in Fig.\,3(a). Since $T^V_K$ represents the sole energy scale characterizing the charge Kondo effect, the conductance as a function of $\Gamma_S/T^V_K$ is universal for different spin voltages [Fig.\,3(b)], provided that $V$ is not strong enough to drive the system out of the Kondo regime and $\Gamma_S$ is not strong enough to extremely suppress the Kondo correlations. The position of the unitary limit in the universal conductance scaling curve gives the constant $\alpha\simeq0.85$ [Fig.\,3(b)]. Thus transport measurements in the hybrid geometry yield a direct determination of the Kondo temperature, which in previous methods can only be determined by further nontrivial analysis. Note that this universal conductance scaling curve
also holds for different negative $U$ as long as it is still in the charge Kondo regime, although we do not present these results here.
\begin{figure}[]
\includegraphics[width=1.0\columnwidth]{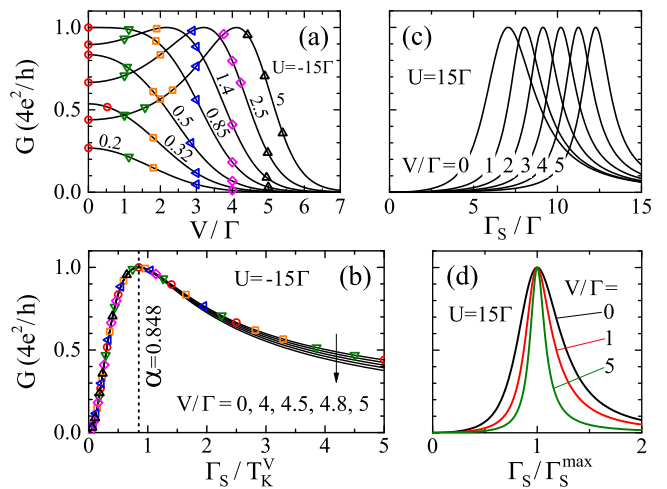}
\caption{(a) Charge conductance $G$ as a function of the spin voltage $V$ for different QD-S couplings $\Gamma_S/T_K$ in the negative-$U$ regime. (b) $G$ as a function of $\Gamma_S$ scaled by the nonequilibrium Kondo temperature $T^V_K$ for different $V$ along the arrow direction in the negative-$U$ regime. Symbols in (a) and (b) correspond to the same conductance data. (c) $G$ as a function of $\Gamma_S$ for different $V$ in the positive-$U$ regime. (d) Same as (c) scaled by $\Gamma_S^\textrm{max}$ the value of QD-S coupling at which $G$ is maximal. $G(V)=G(-V)$ holds in our N-QD-S device.}
\end{figure}

The conductance is, however, strongly suppressed in the positive-$U$ spin Kondo regime even if $\Gamma_S\sim T_K$ at $V=0$. In this regime, the Kondo cotunneling at the N-QD interface and Cooper-pair tunneling at the QD-S interface respectively belong to the odd- and even-occupied subspaces with an energy difference of $\sim\hspace{-0.5mm}\frac{U}{2}$, which cannot mediate a resonant current. In fact, for the device with $U>0$, the maximal conductance approaching the unitary limit can only be achieved near $\Gamma_S\sim\frac{U}{2}$ [Fig.\,3(c)] when the QD-S coupling has compensated the energy difference between the two subspaces and the renormalized pair-tunneling rates at the two interfaces are equal \cite{Tanaka2007,Cuevas2001}. Stronger QD-S couplings are needed to compensate the additional energy difference of $\sim\hspace{-0.5mm}V$ driven by the nonequilibrium spin voltage, thereby shifting rightwards the conductance curves in Fig.\,3(c). Since Kondo correlations are absent in these transport characteristics, a universal conductance scaling for different spin biases does not exist [Fig.\,3(d)]. We would like to further comment that the equilibrium ($V=0$) conductance of the repulsive N-QD-S device was previously obtained by Refs.\,\cite{Tanaka2007,Cuevas2001}. While our results for $V=0$ in Fig.\,3(c) are in quantitative agreement with Ref.\,\cite{Tanaka2007}, the agreement with Ref.\,\cite{Cuevas2001} is only qualitative because the results of Ref.\,\cite{Cuevas2001} were calculated for a finite gap $\Delta$. Nevertheless, this qualitative agreement with the finite gap case implies that the essential features we predict in this paper by considering the large-gap limit could indeed be observed in realistic experiments where the gap is always finite.

\section{IV. Summary}
We have studied the negative-$U$ charge and positive-$U$ spin Kondo effects in a N-QD-S system driven by a spin bias. By mapping it into an equilibrium model solved using the NRG method, the density of states with characteristic Kondo splitting due to the nonequilibrium spin accumulation is obtained at high precision. The novel charge Kondo physics we have revealed, including the additional splitting of the Kondo resonance by superconducting correlations and the direct measurement of the Kondo temperature by the universal conductance scaling, is highly relevant in view of the recent observation of negative-$U$ charge Kondo effect in $\textrm{LaAlO}_3/\textrm{SrTiO}_3$-based QDs \cite{Prawiroatmodjo2017}. Such oxide nanostructures with tunable attractive interaction, individual control of tunnel couplings to different electrodes, and the nanoscale reconfigurability, have paved the way for a new class of investigations of strongly correlated electrons. We hope our paper can stimulate more efforts in this direction.

It is noted that we have not solved the general nonequilibrium Kondo problem. Our method can only tackle the present special nonequilibrium situation where the nonlinear transport is blockaded although the bias voltage is applied out of the linear regime. This is very different from the typical nonequilibrium Kondo transport discussed in Refs.\,\cite{Meir1993,Wingreen1994} where the nonlinear transport channel is open. A precise description of the nonequilibrium Kondo effect in Refs.\,\cite{Meir1993,Wingreen1994} would require to develop truly nonequilibrium methods such as the scattering-states NRG \cite{Anders2008, Schmitt2011} and time-dependent NRG \cite{Anders2005, Eidelstein2012, Nghiem2014, Nghiem2017}. In this context, the numerically exact results for the nonequilibrium Kondo effect in the present paper could be used as valuable nonequilibrium benchmarks for the development of these methods. Previous benchmarks of these methods usually resorted to exact results in the equilibrium or noninteracting cases.

\section{acknowledgments}
We are grateful to Jun-Hong An, Han-Tao Lu, and Yue Ma for inspiring discussions. This work is supported by NSF-China (11574007 and 11504066), NBRP of China (2015CB921102), National Key R and D Program of China (2017YFA0303301), and Innovation-Driven Project of Central South University (2018CX044).

\end{document}